\documentclass[twocolumn,aps,showpacs]{revtex4}
\usepackage{epsfig}
\usepackage[english]{babel}
\usepackage{latexsym}
\usepackage{graphics}
\usepackage{subfigure}
\usepackage{epsfig}
\usepackage{graphicx}
\usepackage{dcolumn}
\usepackage{amsmath}

\newcommand{\be}{\begin{equation}}
\newcommand{\ee}{\end{equation}}
\newcommand{\bey}{\begin{eqnarray}}
\newcommand{\eey}{\end{eqnarray}}
\newcommand{\bw}{\begin{widetext}}
\newcommand{\ew}{\end{widetext}}

\begin{document}

\title{Adiabatic Fidelity for Atom-Molecule Conversion in a Nonlinear
Three-Level $\Lambda$-system }
\author{Shao-Ying Meng$^{1,2}$, Li-Bin Fu$^{1}$, Jie Liu$^{1,}$\footnote{
liu\_jie@iapcm.ac.cn}}

\begin{abstract}
We investigate the dynamics of the population transfer for atom-molecule
three-level $\Lambda$-system on stimulated Raman adiabatic passage(STIRAP).
We find that the adiabatic fidelity for the coherent population
trapping(CPT) state or dark state, as the function of the adiabatic
parameter, approaches to unit in a power law. The power exponent however is
much less than the prediction of linear adiabatic theorem. We further
discuss how to achieve higher adiabatic fidelity for the dark state through
optimizing the external parameters of STIRAP. Our discussions are helpful to
gain higher atom-molecule conversion yield in practical experiments.
\end{abstract}

\affiliation{1. Institute of Applied Physics and Computational Mathematics, P.O.Box 8009
, Beijing 100088, P.R.China \\
2. Graduate School, China Academy of Engineering Physics, P.O.Box 8009-30 ,
Beijing 100088, P.R.China}
\maketitle

In the field of ultracold atomic physics, conversion of an atomic pair to a
molecule by means of photoassociation\cite{kb} or magnetic Feshbach
resonances\cite{tk} is a hot topic both in experiment and in theory.
Photoassociation creates molecules in excited electronic level, while
magneto-association creates molecules in high vibrational quantum state. In
both cases, the resulting molecules are energetically unstable and suffer
from large inelastic loss rate.

One possible scheme to overcome the difficulty is to employ the stimulated
Raman adiabatic passage (STIRAP)\cite{ug,pai,jj}, whose success relies on
the existence of the coherent population trapping(CPT) state, or dark state%
\cite{kwg}. In the traditional atomic $\Lambda $-system, the CPT state
exists when the two-photon resonance condition is satisfied, hence STIRAP
can be straightforwardly implemented by appropriately choosing the laser
frequencies. In parallel to the atomic $\Lambda$-system, the atom-molecule $%
\Lambda$-system also supports a dark state, which facilitates the use of the
STIRAP for creating the ultracold stable molecular state and therefore is a
possible way to achieve the molecular Bose-Einstein condensates(BECs) from
its atomic counterpart\cite{jj}.

Now, a lot of recent theoretical works\cite{jj,hanpu,hyl} have been devoted
to studying the dynamics, approving the adiabatic condition and improving
the conversion efficiency of the atom-molecule coupling model. However,
different from the atomic $\Lambda$- system, the atom-molecule $\Lambda$%
-system is essentially a nonlinear system in which the Hamiltonian is the
functional of instantaneous quantum wavefunctions. More toughly, the system
is no longer invariant under $U(1)$-transformation. With these difficulties,
the adiabatic fidelity that is usually defined by the amplitude of the inner
production between the exact solution and the CPT wavefunction, is not
available, and therefore the quantitative study of the adiabaticity for the
CPT state in the atom-molecule system is still in lack.

In the present paper, we properly define the fidelity for the CPT state in
the nonlinear $\Lambda $-system, and taking advantage of it, we study the
adiabaticity in the atom-molecule conversion system quantitatively. We find
that the adiabatic fidelity for the dark state, as the function of the
adiabatic parameter, approaches to unit in a power law. However, the power
exponents are much less than the prediction from the linear adiabatic
theorem. Then, we further discuss how to optimize the external parameters of
STIRAP process to achieve higher adiabatic fidelity for the dark state.

\begin{figure}[t]
\centering
\rotatebox{0}{\resizebox *{8.0cm}{4.5cm}
{\includegraphics {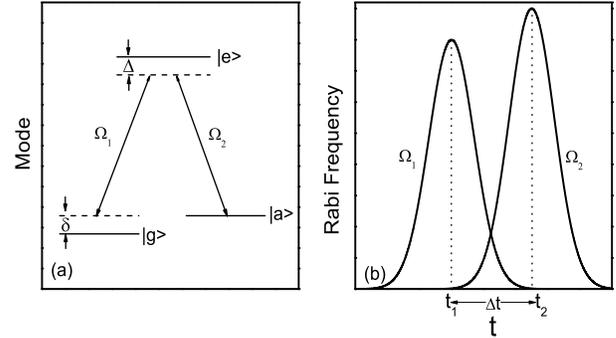}}}
\caption{ (a) Three-level system coupled by two lasers, $\Omega _{1},\Omega
_{2}$ are the Rabi frequencies for the pump and Stokes laser, $\Delta $ and $%
\protect\delta $ are one and two-photon detunings, respectively; (b) Time
dependence of $\Omega _{1},\Omega _{2}$. $t_{1},t_{2}$ are the centers of
the two pulses respectively, and $\Delta t$ is the time delay between the
two pulses. }
\label{scheme}
\end{figure}

Our model is schematically sketched in Fig.\ref{scheme}(a). The initial
state $|a\rangle$(atomic state) and the intermediate state $|e\rangle$%
(excited molecular state) are coupled by pump laser with Rabi frequency $%
\Omega_{2}$, while state $|e\rangle$ and the target state $|g\rangle$(ground
molecular state) are coupled by stokes laser with Rabi frequency $\Omega_{1}$%
. The frequencies of the applied lasers are expressed in terms of the
single- and two-photon detunings $\Delta$ and $\delta$, respectively.
Without loss of generality, we assume that the Rabi frequencies $%
\Omega_{1,2} $ are real and positive. Under the two-photon resonance
condition, the Hamiltonian in second quantized form reads,

\begin{equation}
H_{am}=-\hbar \Delta \hat{\psi _{e}}^{\dag }\hat{\psi _{e}}+\frac{\hbar }{2}
\left( -\Omega _{2}\hat{\psi _{e}}^{\dag }\hat{\psi _{a}}\hat{\psi _{a}}
+\Omega _{1}\hat{\psi _{e}}^{\dag }\hat{\psi _{g}}+h.c.\right) .  \label{a}
\end{equation}
where $\hat{\psi _{i}}$ and $\hat{\psi _{i}}^{\dag }$ are the annihilation
and creation operators for state $|i\rangle $, respectively. Under the
mean-field approximation, i.e., $\hat{\psi _{i}}$ and $\hat{\psi _{i}}^{\dag
}$ are replaced by $c-$number $\psi _{i}$ and $\psi _{i}^{\ast }$, the
Schr\"odinger equation are,
\begin{subequations}
\begin{equation}
i\dot{\psi _{a}}=-\Omega _{2}\psi _{a}^{\ast }\psi _{e},
\end{equation}
\begin{equation}
i\dot{\psi _{e}}=-\Delta \psi _{e}-\frac{\Omega _{2}}{2}\psi _{a}^{2}+\frac{%
\Omega _{1}}{2}\psi _{g},
\end{equation}
\begin{equation}
i\dot{\psi _{g}}=\frac{\Omega _{1}}{2}\psi _{e}.
\end{equation}
In the above model, the nonlinear collisions between particles are
neglected, so the only nonlinearity comes from the fact that it takes two
atoms to form a molecule. Mathematically, we see that the Hamiltonian in the
above Schr\"odinger equation is the functional of the instantaneous
wavefunction as well as its conjugate.

The adiabatic theory for nonlinear quantum systems has been set up recently
in paper\cite{jl}, where new adiabatic conditions and adiabatic invariants
are given. However, these discussions are restricted to systems that have $%
U(1)$ invariance. For the atom-molecule $\Lambda $-system, because the
Hamiltonian is the functional of both the wavefunction and its conjugate,
the $U(1)$-invariance is broken. Instead, the system is invariant under the
following transformation,
\end{subequations}
\begin{equation}
U(\phi )=e^{i\Theta (\phi )},\Theta (\phi )=\left(
\begin{array}{ccc}
\phi  & 0 & 0 \\
0 & 2\phi  & 0 \\
0 & 0 & 2\phi
\end{array}%
\right) ,  \label{uu}
\end{equation}%
Under this transformation, $|\psi \rangle =\left( \psi _{a},\psi
_{e},\psi _{g}\right) ^{T}\rightarrow $ $|\psi ^{^{\prime }}\rangle
=U(\phi )|\psi \rangle =\left( \psi _{a}e^{i\phi },\psi
_{e}e^{i2\phi },\psi _{g}e^{i2\phi }\right) ^{T}.$ In fact, when the
diagonal terms in the above matrix are identical,  the
transformation $U(\phi )$ degenerates to the $U(1)$ transformation.

This new kind of  symmetry allows the following stationary states
with chemical potential $\mu $,
\begin{equation}
\psi _{a}=\left\vert \psi _{a}\right\vert e^{i\theta _{a}}e^{-i\mu t},~~\psi
_{e,g}=\left\vert \psi _{e,g}\right\vert e^{i\theta _{e,g}}e^{-i2\mu t},
\label{ua}
\end{equation}%
where
\begin{equation}
2\theta _{a}-\theta _{e}=constant,~~\theta _{g}-\theta _{e}=constant.
\label{ub}
\end{equation}%
Under the normalized condition $\left\vert \psi _{a}\right\vert
^{2}+2\left\vert \psi _{e}\right\vert ^{2}+2\left\vert \psi _{g}\right\vert
^{2}=1$, it is easy to show, as in the atomic counterpart, the atom-molecule
$\Lambda -$ system supports a CPT eigenstate \cite{hanpu,hyl} with zero
eigenvalue.
\begin{equation}
|CPT\rangle =\left( \frac{(\Omega _{1}\Omega _{eff}^{nl}-\Omega _{1}^{2})^{%
\frac{1}{2}}}{2\Omega _{2}},0,\frac{\Omega _{eff}^{nl}-\Omega _{1}}{4\Omega
_{2}}\right) ^{T},  \label{5}
\end{equation}%
where $\Omega _{eff}^{nl}=\sqrt{\Omega _{1}^{2}+8\Omega _{2}^{2}}$.

When the Rabi laser pulses are ramped up adiabatically, i.e., $\Omega_{1,2}$
vary in time slowly, an state that is initially prepared as the CPT state is
expected to close to the instantaneous CPT state during the whole process.
The problem is how close is the above adiabatic approximation. To clarify
the above question and formulate it quantitatively, we introduce two
physical quantities, namely the adiabatic fidelity and the adiabatic
parameter.

For the linear system, adiabatic fidelity is introduced as $%
F_{ad}=\left\vert\langle\psi(t)|\psi_{ad}\rangle\right\vert ^{2}$, where $%
|\psi_{ad}\rangle$ and $|\psi(t)\rangle$ are defined as the adiabatic
approximate solution and the real one. The adiabatic fidelity approaches to
unit in a power law of the adiabatic parameter\cite{pa}, i.e., $%
1-F_{ad}\sim\epsilon^{2}$. Here, the adiabatic parameter $\epsilon$ is the
ratio between the rate of energy's change and level space. For linear
quantum systems, evaluating the adiabatic fidelity gives an good estimation
on how close the real solution is to the adiabatic approximate solution\cite%
{liufu}.

For the atom-molecule nonlinear system, the traditional definition of
fidelity is no longer suitable because the system is not invariant under $%
U(1)$-transformation. We  need to define new fidelity for such
system. For convenience, we denote the fidelity of two states $|\psi
_{1}\rangle $ and $|\psi _{2}\rangle $ as $F^{am}(|\psi _{1}\rangle
,|\psi _{2}\rangle )$  that should satisfy  $F^{am}(|\psi \rangle
,U(\phi )|\psi \rangle )=1$  for any $\phi$. With this
consideration, we define the adiabatic fidelity as $F^{am}(|\psi
_{1}\rangle ,|\psi _{2}\rangle )=\left\vert \left\langle
\overline{\psi _{1}}\right\vert
\overline{\psi _{2}}\rangle \right\vert ^{2}$, where $|\overline{\xi }%
\rangle =\left( \xi _{a}^{2}/|\xi _{a}|,\sqrt{2}\xi _{e},\sqrt{2}\xi
_{g}\right) ^{T}$ is the rescaled wavefunction of $|\xi \rangle =\left( \xi
_{a},\xi _{e},\xi _{g}\right) ^{T}$. One can prove that the definition
satisfies the above conditions and the others for fidelity definition \cite%
{fide}. Because we only concern the adiabatic evolution of the CPT
state throughout, we denote the adiabatic fidelity as
$F^{am}=\left\vert \langle \overline{\psi (t)}|\overline{CPT}\rangle
\right\vert ^{2}$ where $|\psi (t)\rangle =\left( \psi _{a},\psi
_{e},\psi _{g}\right) ^{T}$ is the exact solution of the
Schr\"{o}dinger equation.

For the atom-molecule three-level $\Lambda $-system, owing to the invalid of
the concept of an orthogonal set of energy eigenstates and the linear
superposition principle involving these states, the adiabatic parameter has
nothing to do with the energy level spacing\cite{jl}. Moreover, noticing
that the eigenstates correspond to extremum points or fixed points of the
system energy, the fundamental frequencies of periodic orbits around the
fixed points serves as the adiabatic parameter. These frequencies can be
evaluated by linearizing Eq.(2) about the fixed points and are identical to
the Bogolubov excitation spectrum of the corresponding eigenstate, as is
demonstrated in \cite{jl,hanpu}, then the adiabatic parameter is expressed
as,
\begin{equation}
\epsilon ^{am}=\left\vert \frac{\dot{\Omega}_{1}\Omega _{2}-\Omega _{1}\dot{%
\Omega _{2}}}{\Omega _{1}+\Omega _{eff}^{nl}}\right\vert \frac{1}{\Omega
_{eff}^{nl}\Omega _{1}/2} ~.  \label{25}
\end{equation}
For the atom-molecule $\Lambda$-system, we introduce
\begin{equation}
\Omega_{1,2}=\Omega_{1,2}^{^{\prime }}e^{-(t-t_{1,2})^{2}}.  \label{26}
\end{equation}

The property of the system is related to three parameters, i.e., the
amplitude of the two pulses $\Omega_{1,2}^{^{\prime }}$ and the time delay $%
\Delta t$. So, our discussions are divided in two cases: the two pulses
having equal amplitudes and unequal amplitudes. In each case, the dependence
on different time delays will also be studied.

Firstly, for convenience, we consider the two pulses having equal
amplitudes, i.e., $\Omega_{1}^{^{\prime }}=\Omega_{2}^{^{\prime
}}=\Omega_{0} $. Substituting Eq.(\ref{26}) into Eq.(\ref{25}), we obtain
the adiabatic parameter of the atom-molecule system:
\begin{equation}
\epsilon ^{am}\sim \frac{1}{\Omega _{0}}.
\end{equation}

\begin{figure}[t]
\centering
\rotatebox{0}{\resizebox *{8.0cm}{5.0cm}
{\includegraphics {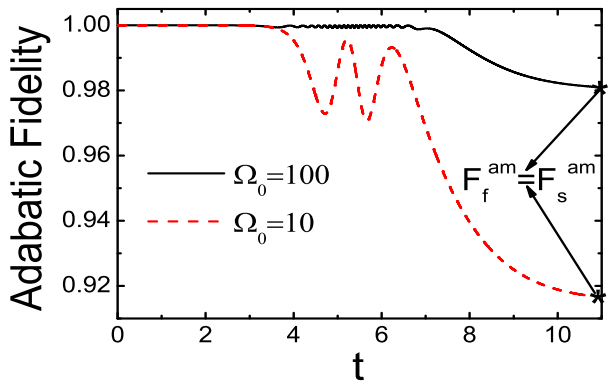}}}
\caption{(Color online)Time evolution of the adiabatic fidelity for the
atom-molecule three-level system with $\Delta =0,t_{1}=5.0,t_{2}=6.0,\Delta
t=1.0,\Omega _{0}=10,100$.}
\label{ft}
\end{figure}
In the STIRAP, it is required that $t_{1}<t_{2}$ and only level $|a\rangle $
is populated initially, then the system evolve under $H_{am}$.

For easy to understand the dynamics of the population transfer, in figure %
\ref{ft}, we show a typical change of the adiabatic fidelity with
time for the atom-molecule case with $\Delta t=1.0$. Here and
henceforth, the exact solution $\Psi(t)>$ is obtained from directly
solving Eq.(2) numerically using the 4-5 order Runge-Kutta
algorithm. We only focus on the final and
the smallest adiabatic fidelity which is marked with $F_{f}^{am}$ and $%
F_{s}^{am}$(which points are labelled by $\ast $ in figure 2) to understand
the whole property of the system, since $F_{s}^{am}$ denotes the maximal
deviation from CPT state in the whole process and $F_{f}^{am}$ denotes the
final conversion efficiency. The bigger $F_{s}^{am}$ is, the better the
adiabaticity is. And the bigger $F_{f}^{am}$ is, the higher the conversion
efficiency is. From this figure, we can see that $F_{f}^{am}=F_{s}^{am}$ and
$F_{f}^{am}(\Omega_{0}=100)>F_{f}^{am}(\Omega_{0}=10)$. Indeed, we are
interested in the relationship between $F_{f,s}^{am}$ and the amplitude $%
\Omega_{0}$, and moreover we can further find the relationship between the
adiabatic fidelity and the adiabatic parameter.

\begin{figure}[t]
\centering
\rotatebox{0}{\resizebox *{8.0cm}{5.0cm}
{\includegraphics {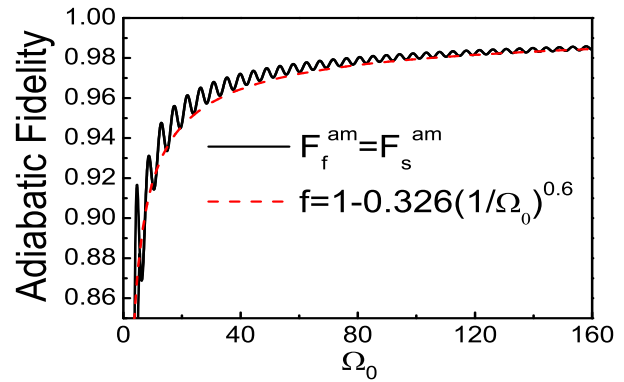}}}
\caption{(Color online)Final and smallest adiabatic fidelity for the
atom-molecule three-level system as a function of the amplitude $\Omega_{0}$
with $\Delta=0,t_{1}=5.0,t_{2}=6.0,\Delta t=1.0.$}
\label{ffa}
\end{figure}

Figure \ref{ffa} shows the final adiabatic fidelity $F_{f}^{am}$ and the
smallest adiabatic fidelity $F_{s}^{am}$ as a function of $\Omega _{0}$ for
the atom-molecule system with $\Delta t=1.0$. From this figure, one can see
that the larger the amplitude $\Omega _{0}$ is, the bigger $F_{f,s}^{am}$
is, which is conformed by figure 2(where $F_{f}^{am}(%
\Omega_{0}=100)>F_{f}^{am}(\Omega_{0}=10)$). We obtain the asymptotic
behaviors of the lower bounds of $F_{f}^{am}$ and $F_{s}^{am}$ which can be
expressed as
\begin{equation}
1-F_{f}^{am}=1-F_{s}^{am}=0.326\left( \frac{1}{\Omega _{0}}\right)
^{0.6}\sim \left( {\epsilon ^{am}}\right) ^{0.6}.  \label{a2}
\end{equation}

As in the linear system, there exists a power law relationship between the
adiabatic fidelity and the adiabatic parameter in the atom-molecule
conversion system, but the exponent is 0.6 rather than 2. Here, the power
exponent is not universal, it depends on the time delay $\Delta t$. For
example, when $\Delta t=0.8$, it is 0.5, $\Delta t=1.2$, it is 1.3.

In the above, we have discussed the situation when the two pulses having
equal amplitudes $\Omega_{0}$ with $\Delta t=1.0$, and find that the larger
the amplitude $\Omega _{0}$ is, the better the adiabaticity is, and hance
the higher the conversion efficiency is. That is to say, we can optimize
conversion efficiency by increasing the amplitude of the pulses $\Omega_{0}$.

\begin{figure}[t]
\centering
\rotatebox{0}{\resizebox *{8.0cm}{5.0cm}
{\includegraphics {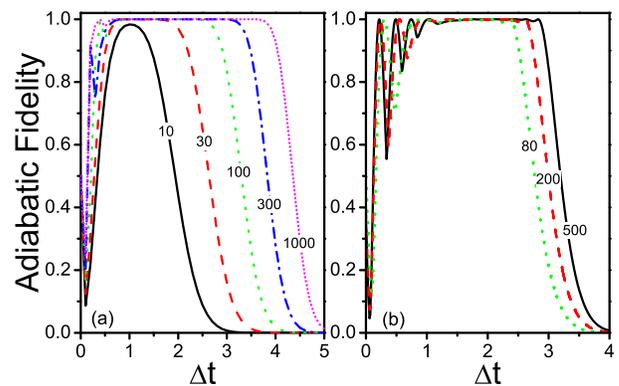}}}
\caption{(Color online)Final adiabatic fidelity of atom-molecule three-level
system plotted against the pulse delay $\Delta t$ for: (a) different $\Omega
_{0}$(denoted on the respective curves with numbers); (b) different $%
\Omega_{2}^{^{\prime }}$ (denoted on the respective curves with numbers) at $%
\Omega_{1}^{^{\prime }}=10$}
\label{ff1}
\end{figure}

How the conversion efficiency changes with the time delay is also
interesting because it is most concerned in practical experiments. In figure %
\ref{ff1}(a), the final adiabatic fidelity $F_{f}^{am}$ is plotted as a
function of the pulse delay $\Delta t$ for several different values of $%
\Omega_{0}$ in the atom-molecule system. We can see that, as the delay $%
\Delta t$ increases, the final adiabatic fidelity increases firstly, then
reaches a steady stage depending on the amplitude $\Omega_{0}$, finally
decreases beyond certain values of $\Delta t$, which is similar to the
result in atomic system in article\cite{pai}. So, for the atom-molecule
system, there is a optimum segment of time delay $\Delta t$, which increases
as the amplitude $\Omega_{0}$ grows, making the conversion efficiency to the
be optimal.

Now, we turn to consider the two pulses having unequal amplitudes, and study
the conversion efficiency and the adiabatic property, further find how to
optimize conversion efficiency.
\begin{figure}[tbh]
\centering
\rotatebox{0}{\resizebox *{8.0cm}{5.0cm}
{\includegraphics {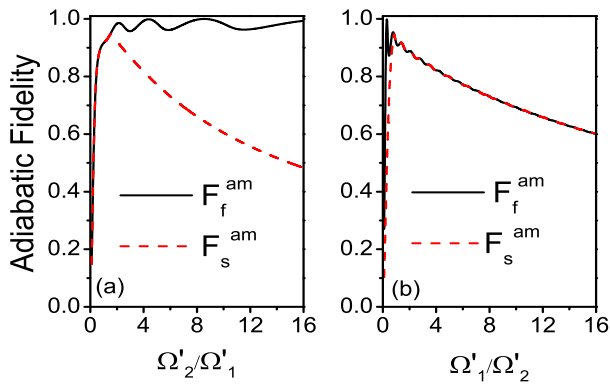}}}
\caption{(Color online)Final and smallest adiabatic fidelity for
atom-molecule three-level system (a) as a function of $\Omega _{2}^{^{\prime
}}/\Omega _{1}^{^{\prime }}$ with fixed $\Omega _{1}^{^{\prime }}=10$ and
(b) as a fuction of $\Omega _{1}^{^{\prime }}/\Omega _{2}^{^{\prime }}$ with
fixed $\Omega _{2}^{^{\prime }}=10$, $\Delta =0,t_{1}=5.0,t_{2}=6.0,\Delta
t=1.0.$}
\label{ffau1}
\end{figure}

Figure \ref{ffau1} shows $F_{f}^{am}$ and $F_{s}^{am}$ as a function of $%
\Omega_{2}^{^{\prime }}/\Omega_{1}^{^{\prime }}$ and $\Omega _{1}^{^{\prime
}}/\Omega _{2}^{^{\prime }}$ with $\Delta t=1.0$. In figure \ref{ffau1}(a), $%
\Omega _{1}^{^{\prime }}$ is fixed, as $\Omega_{2}^{^{\prime }}$ increases,
we can see that both $F_{f}^{am}$ and $F_{s}^{am}$ increase until reach a
critical point (at about $\Omega_{2}^{^{\prime }}=15$ in figure \ref{ffau1}%
(a)), $F_{s}^{am}$ begins to decrease while $F_{f}^{am}$ stays close at 1
with vibration, which means the adiabaticity of the system is weakened as $%
\Omega_{2}^{^{\prime}}$ increases, but the process can still realize
completely population transfer. The physics behind this is that the
effective coupling between $|a\rangle$ and $|e\rangle$ scaled as $%
\Omega_{2}\psi_{a}$ is weakened by decreasing atom population on $|a\rangle$%
, hence a higher pump Rabi frequency can remedy this kind of weakening, and
enhance the two-photon process. So, we can improve the conversion efficiency
by enlarging the amplitude of the pump laser $\Omega_{2}^{^{\prime }}$ with
the optimal match between $\Omega_{2}^{^{\prime }}$ and $\Omega_{1}^{^{%
\prime }}$. However, we cannot increase the conversion efficiency through
increasing the amplitude of the stoke laser $\Omega_{1}^{^{\prime }}$,
because both $F_{f}^{am}$ and $F_{s}^{am}$ decrease after the critical
points, as is shown in figure \ref{ffau1}(b). Therefore, in the following
discussion, we only consider how the conversion efficiency changes with the
time delay when $\Omega_{1}^{^{\prime }}$ is fixed and $\Omega_{2}^{^{\prime
}}$ is changed.

In figure \ref{ff1}(b), the final adiabatic fidelity $F_{f}^{am}$ is plotted
as a function of the pulse delay $\Delta t$ for several different values of $%
\Omega_{2}^{^{\prime }}$ with $\Omega_{1}^{^{\prime }}=10$. We find that, as
the case of equal amplitudes, there are also a segment and some preferable
points of time delay $\Delta t$ making the conversion efficiency of the
atom-molecule system to be the optimal. Moreover the segment increases as
the the amplitude of the pump laser $\Omega_{2}^{^{\prime }}$ grows.

In conclusion, we study the adiabatic fidelity for the the coherent
population trapping(CPT) state in the atom-molecule three-level $\Lambda$%
-system on stimulated Raman adiabatic passage(STIRAP) quantitatively, and
discuss how to achieve higher adiabatic fidelity for the dark state through
optimizing the external parameters of STIRAP. Our discussions are helpful to
achieve higher atom-molecule conversion rate in practical experiments.

This work is supported by National Natural Science Foundation of China
(No.10474008,10604009), the National Fundamental Research Programme of China
under Grant No. 2006CB921400,2007CB814800.


\begin{thebibliography}{99}
\bibitem{kb} K. Bergmann, H. Theuer, and B. W. Shore, Rev. Mod. Phys.
\textbf{70} , 1003 (1998).

\bibitem{tk} E. Timmermans, et all. Phys. Rep. \textbf{315}, 199 (1999);
Thorsten K\"ohler, Krzysztof G\'oral and Paul S. Julienne. Rev. Mod. Phys.
\textbf{78(4)}, 1311 (2006), and reference therein.

\bibitem{ug} U. Gaubatz, et al, Chem. Phys. Lett. \textbf{149}, 463 (1988);
J.R.Kuklinski, et al. Phys. Rev. A. \textbf{40} 1989; N. V. Vitanov, et al,
Annu. Rev. Phys. Chem. \textbf{52}, 763 (2001).

\bibitem{pai} P. A. Ivano, N. V. Vitanov, and K. Bergmann, Phys. Rev. A.
\textbf{70} , 063409 (2004); \textbf{72}, 053412 (2005).

\bibitem{jj} J. J. Hope, M. K. Olsen, and L. I. Plimak, Phys. Rev. A.
\textbf{63}, 043603 (2001); M. Mackie, R. Kowalski, and J. Javanainen, Phys.
Rev. Lett. \textbf{84}, (2000); M. Mackie, et al, Phys. Rev. A. \textbf{70},
013614 (2004); M. Mackie, A. Collin, and J. Javanainen, Phys. Rev. A.
\textbf{71}, 017601 (2005); P. D. Drummond, et al, Phys. Rev. A. \textbf{65}%
, 063619(2002); \textbf{71}, 017602 (2005).

\bibitem{kwg} K. Winkler, et al, Phys. Rev. Lett. \textbf{95}, 063202
(2005); S. Moal, et al, Phys. Rev. Lett. \textbf{96}, 023203(2006).

\bibitem{hanpu} Han Pu, Peter Maenner, Weiping Zhang, and Hong Y.Ling. Phys.
Rev. Lett. \textbf{98}, 050406 (2007);

\bibitem{hyl} Hong Y. Ling, Han Pu, and Brian Seaman, Phys. Rev. Lett.
\textbf{93}, 250403 (2004); Hong Y. Ling, Peter Maenner, Weiping Zhang, and
Han Pu. Phys. Rev. A. \textbf{75}, 033615 (2007).

\bibitem{jl} Jie Liu, Biao Wu, and Qian Niu, Phys. Rev. Lett. \textbf{90},
170404 (2000).

\bibitem{pa} P. Ehrenfest, Ann. Phys. (Berlin) \textbf{51}, 327 (1916); M.
Born and V. Fock, Z. Phys. \textbf{51}, 165 (1928); L. D. Landau,
Zeitschrift, \textbf{2}, 46 (1932); C. Zener, Proc. R. Soc. A \textbf{137},
696 (1932).

\bibitem{liufu} Jie Liu, Li-Bin Fu, \emph{Singularity of Berry Connections
Inhibits the Accuracy of Adiabatic Approximation}, arXiv:quant-ph/0612088
(2006), to appear in Phys. Lett. A.

\bibitem{fide} M. A. Nielsen and I. L. Chuang, Quantum Computation and
Quantum Information (Cambridge University Press, Cambridge, England, 2000),
pp. 399--424; Jing-Ling Chen, Libin Fu, Abraham A. Ungar, and Xian-Geng
Zhao, Phys. Rev. A 65, 054304 (2002); Jing-Ling Chen, Libin Fu, Abraham A.
Ungar, and Xian-Geng Zhao, Phys. Rev. A 65, 024303 (2002).
\end{thebibliography}
\end{document}